\documentclass[letters,useAMS,usenatbib,usegraphicx]{mn2e}

\newcommand{\kep}{{\em Kepler}}
\newcommand{\kasc}{{\sc kasc}}
\newcommand{\kepmi}{{\em Kepler Mission}}
\newcommand{\msol}{\ensuremath{\rm{M}_{\odot}}}
\newcommand{\menv}{\ensuremath{\rm{M}_\mathrm{env}}}
\newcommand{\teff}{\ensuremath{T_{\rm{eff}}}}
\newcommand{\logg}{\ensuremath{\log g}}

\newcommand{\lheh}{\ensuremath{\log \left(N_{\mathrm{He}}/N_{\mathrm{H}}\right)}}

\newcommand{\twom}{{\sc 2mass}}
\newcommand{\uHz}{\ensuremath{\umu{\rm{Hz}}}}
\newcommand{\uma}{\ensuremath{\umu{\rm{ma}}}}
\newcommand{\kms}{\rm{km}\,\rm{s}\ensuremath{^{-1}}}
\newcommand{\kic}{{\sc kic}}

\newcommand{\forb}{\ensuremath{f_{\rm{orb}}}}
\newcommand{\flc}{\ensuremath{f_{\rm{lc}}}}

\newcommand{\fnyq}{\ensuremath{f_{\rm{Nyq}}}}
\newcommand{\sig}{\ensuremath{\sigma}}
\newcommand{\target}{2M1938+4603}
\newcommand{\kictarg}{\kic\,9472174}
\newcommand{\ueq}[3]{\ensuremath{#1}\,=\,{#2}\,{\rm{#3}}}
\newcommand{\eueq}[4]{\ensuremath{#1}\,=\,{#2}\,\ensuremath{\pm}{#3}\,{\rm{#4}}}

\title[2M1938+4603: An eclipsing sdBV+dM observed with {\em Kepler}]
{2M1938+4603: A rich, multimode pulsating sdB star\\
with an eclipsing dM companion observed with {\em Kepler\/}}
\author[R.~H.~{\O}stensen et al.]
       {R.~H.~{\O}stensen,$^1$\thanks{E-mail: roy@ster.kuleuven.be}
        E.~M.~Green,$^{2}$                
        S.~Bloemen,$^{1}$                 
        T.~R.~Marsh,$^{3}$                
        J.~B.~Laird,$^{2}$ \newauthor     
        M.~Morris,$^{2}$                  
        E.~Moriyama,$^{2}$                
        R.~Oreiro,$^{4}$
        M.~D.~Reed,$^{5}$
        S.~D.~Kawaler,$^{6}$
        C.~Aerts,$^{1,7}$ \newauthor
        M.~Vu\v{c}kovi\'{c},$^{8}$
        P.~Degroote,$^{1}$
        J.~H.~Telting,$^{9}$
        H.~Kjeldsen,$^{10}$
        R.~L.~Gilliland,$^{11}$ \newauthor
        J.~Christensen-Dalsgaard,$^{10}$
        W.~J.~Borucki$^{12}$ and
        D.~Koch$^{12}$\\
        $^1$Instituut voor Sterrenkunde, K.~U.~Leuven, Celestijnenlaan 200D,
        3001 Leuven, Belgium\\
        $^2$Steward Observatory, University of Arizona, 933 N.~Cherry Ave.,
        Tucson, AZ 85721, USA \\
        $^3$Department of Physics, University of Warwick,
        Coventry CV4 7AL, UK \\
        $^4$Instituto de Astrof\'isica de Andaluc\'ia,
        Glorieta de la Astronom\'ia s/n, 18008 Granada, Spain \\
        $^5$Department of Physics, Astronomy, and Materials Science,
        Missouri State University, Springfield, MO 65804, USA \\
        $^6$Department of Physics and Astronomy, Iowa State University,
        Ames, IA 50011, USA \\
        $^7$Department of Astrophysics, IMAPP, Radboud University
        Nijmegen, 6500 GL Nijmegen, The Netherlands \\
        $^8$European Southern Observatory, Alonso de C\'ordova 3107,
        Vitacura, Casilla 19001, Santiago, Chile \\
        $^9$Nordic Optical Telescope, 38700 Santa Cruz de La Palma, Spain \\
        $^{10}$Department of Physics and Astronomy, Aarhus University,
        8000 Aarhus C, Denmark \\
        $^{11}$Space Telescope Science Institute, 3700 San Martin Drive,
        Baltimore, MD 21218, USA \\
        $^{12}$NASA Ames Research Center, MS 244-30, Moffett Field, CA 94035,
        USA
}
\begin{document}

\date{Released 2010 Xxxxx XX}
\pagerange{\pageref{firstpage}--\pageref{lastpage}} \pubyear{2010}
\maketitle
\label{firstpage}

\begin{abstract}
2M1938+4603 (\kic\,9472174)
displays a spectacular light curve dominated by a strong
reflection effect and rather shallow, grazing eclipses.
The orbital period is 0.126 days, the second longest period yet
found for an eclipsing sdB+dM, but still close to
the minimum 0.1-d period among such systems. The
phase-folded light curve was used to detrend the orbital effects from
the dataset, and the resulting amplitude spectrum shows a rich
collection of pulsation peaks spanning frequencies from $\sim$50 to
4500\,\uHz. The presence of a complex pulsation spectrum in both the
$p$-mode and the $g$-mode regions has never been seen before in
a compact pulsator.

Eclipsing sdB+dM stars are very rare, with only seven systems known and
only one with a pulsating primary. Pulsating stars in eclipsing binaries
are especially important since they permit masses derived from seismological
model fits to be cross checked with orbital mass constraints.
We present a first analysis of this star based on the \kep\ 9.7-day
commissioning light curve and extensive ground-based photometry and
spectroscopy that allow us to set useful bounds on the system parameters.
We derive a radial-velocity amplitude \eueq{K_1}{65.7}{0.6}{\kms},
inclination angle \eueq{i}{69.45}{0.20}{$^\circ$}, and find that the
masses of the components are \eueq{M_1}{0.48}{0.03}{\msol} and
\eueq{M_2}{0.12}{0.01}{\msol}.
\end{abstract}

\begin{keywords}
   subdwarfs --
   binaries: close -- binaries: eclipsing --
   stars: variables: general --
   stars: individual: 2M1938+4603
\end{keywords}

\section{Introduction}

The subdwarf B (sdB) stars are known to be core helium burning stars
with extremely thin (\menv\,$\leq$\,0.01\,\msol)
inert hydrogen dominated envelopes \citep{heber86}.
This places them on an extension to the classical horizontal branch,
known as the extreme horizontal branch (EHB).
In order to reach this configuration, almost the entire
envelope must be stripped off close to the tip of the red giant branch.
There are several binary scenarios that are capable of accomplishing
this, such as common-envelope ejection (CEE), stable Roche-lobe overflow and
merger of two helium-core white dwarfs \citep{han02}.
A small minority of sdB stars in the field are found to have close M-dwarf
companions \citep[see][for recent results and a review]{for10}, and
seven such systems are eclipsing.
The eclipsing systems have been monitored over long time-bases in order to
detect low-mass companions from precise
measurements of the eclipse timings with the O--C method
\citep{kilkenny03_HWVir,lee09}.

One of the eclipsing sdB+dM systems, NY\,Vir, has a pulsating primary
of the V361\,Hya class \citep{kilkenny98}.
The V361\,Hya stars were discovered by
\citet{kilkenny97} and are characterised by rapid pulsations,
typically in the period range between two and five minutes.
They are known to be pressure
($p$-)mode pulsators excited by the $\kappa$ mechanism, driven primarily
by an iron opacity bump in the envelope \citep{charpinet97}.
A second class of sdB pulsators was reported by \citet{green03}.
These stars, now known as V1093\,Her stars, show pulsations with 
much longer periods ($\sim$1\,h) than the V361\,Hya stars,
and their temperatures are lower. These pulsations can be described
in terms of gravity ($g$-)modes excited by the same $\kappa$ mechanism
\citep{fontaine03}.  With the discovery of long-period pulsations in
a known rapid pulsator, DW\,Lyn, \citet{schuh06} established
the existence of hybrid sdB pulsators.
A broad review of hot subdwarf stars in general can be found in
\citet{heber09}, and a review of asteroseismology and evolution of
the EHB stars can be found in \citet{ostensen09}. For the most recent
pulsator discoveries see \citet{sdbnot}.

The \kep\ spacecraft was launched in March 2009, aiming to find
Earth-sized planets from photometric observations of a 105 square
degree field \citep{borucki10}.
\kep~is also ideally suited for asteroseismological studies,
and the Kepler Asteroseismic Science Consortium
\citep[\kasc,][]{gilliland10a} manages this important aspect of the mission.
The methods
with which the compact pulsator candidates were selected, together with
analysis of the first half of the survey phase, are presented in
\citet[][Paper\,{\sc i}]{kepI}.
The first results on a V361\,Hya star in the \kep\ field are presented
by \citet{kawaler10a}, the first results on V1093\,Her and DW\,Lyn pulsators
are presented by \citet{reed10a}, and results on two V1093\,Her pulsators in
sdB+dM reflection binaries are discussed in \citet{kawaler10b}.
Further analysis of \kep\ data on sdB stars are found in
\citet{VanGrootel10}, who present the first detailed asteroseismic
analysis of a V1093\,Her star, and in \citet{bloemen10}, who present
a detailed analysis of the extraordinary binary light curve of the sdB+WD 
star KPD\,1946+4340.

In this letter we present \target~(\kictarg),
an object first identified as an sdB star in the \kep\ field during a
survey of blue targets selected from \twom\ photometry \citep{twomass}.
Follow-up photometry showed the presence of a reflection
effect with shallow primary and secondary eclipses (Fig.~\ref{fig:lc}).
The amplitude of the reflection effect is very comparable to those
observed in HW\,Vir and NY\,Vir, but the eclipses are much more shallow.
After detrending the \kep\ light curve, we clearly detect low-level
pulsations.

\section{Observations}

The discovery of the strong reflection effect with grazing eclipses
was made by two of us (JBL \& MM) during a photometric run in
June 2008.
Here we present only the eclipse timings from
the ground-based photometry, as the pulsations are too complex and have
too low amplitudes to be significant in those light curves.
The times of thirteen primary eclipses
collected between June 2008 and May 2010
(Table~\ref{tbl:ephem1}) give the following ephemeris:
\begin{eqnarray*}
T_{0} & = &  2454640.864162 \pm 0.000058\,{\mathrm d} \\
P     & = &   0.125765300 \pm 0.000000021\,{\mathrm d}.
\end{eqnarray*}
We downloaded the \kep\ Q0 light curve from the \kasc~archive,
corrected the raw time-stamps to Barycentric Julian Date (BJD)
according to the instructions \citep{kepler_dr02},
and
made minor corrections to the raw fluxes to take out trends on
time-scales longer than a day.
The 77 consecutive eclipse times measured from the \kep\ photometry
are listed in Table~\ref{tbl:ephem2}, and match the ground-based ephemeris 
well within the errors.

\begin{figure}
\centering
\includegraphics[width=\hsize,clip]{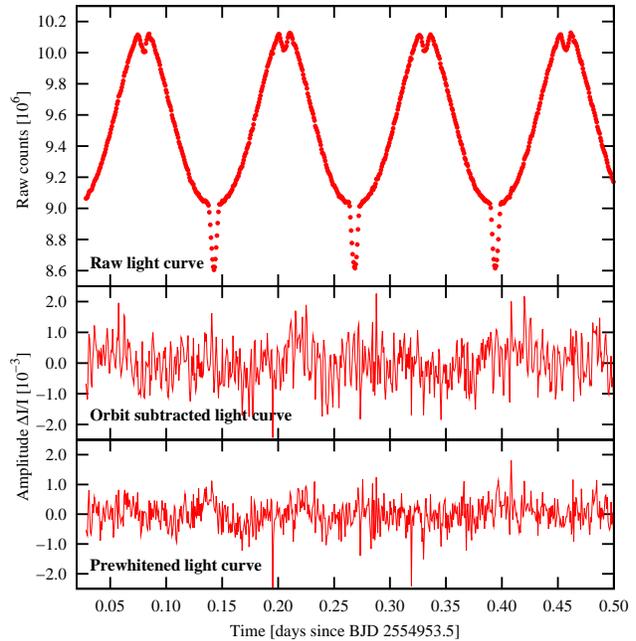}
   \caption{The first half day of the \kep\ light curve. The upper panel
            shows the raw data, which is dominated by the strong reflection
            effect, with both primary and secondary eclipses clearly visible.
            The middle panel shows the same chunk after detrending with the
            light curve folded on $P$, and the bottom panel shows
            the same after prewhitening the 55 most significant frequencies.
           }
      \label{fig:lc}
\end{figure}

\begin{figure*}
\centering
\includegraphics[width=\textwidth,clip]{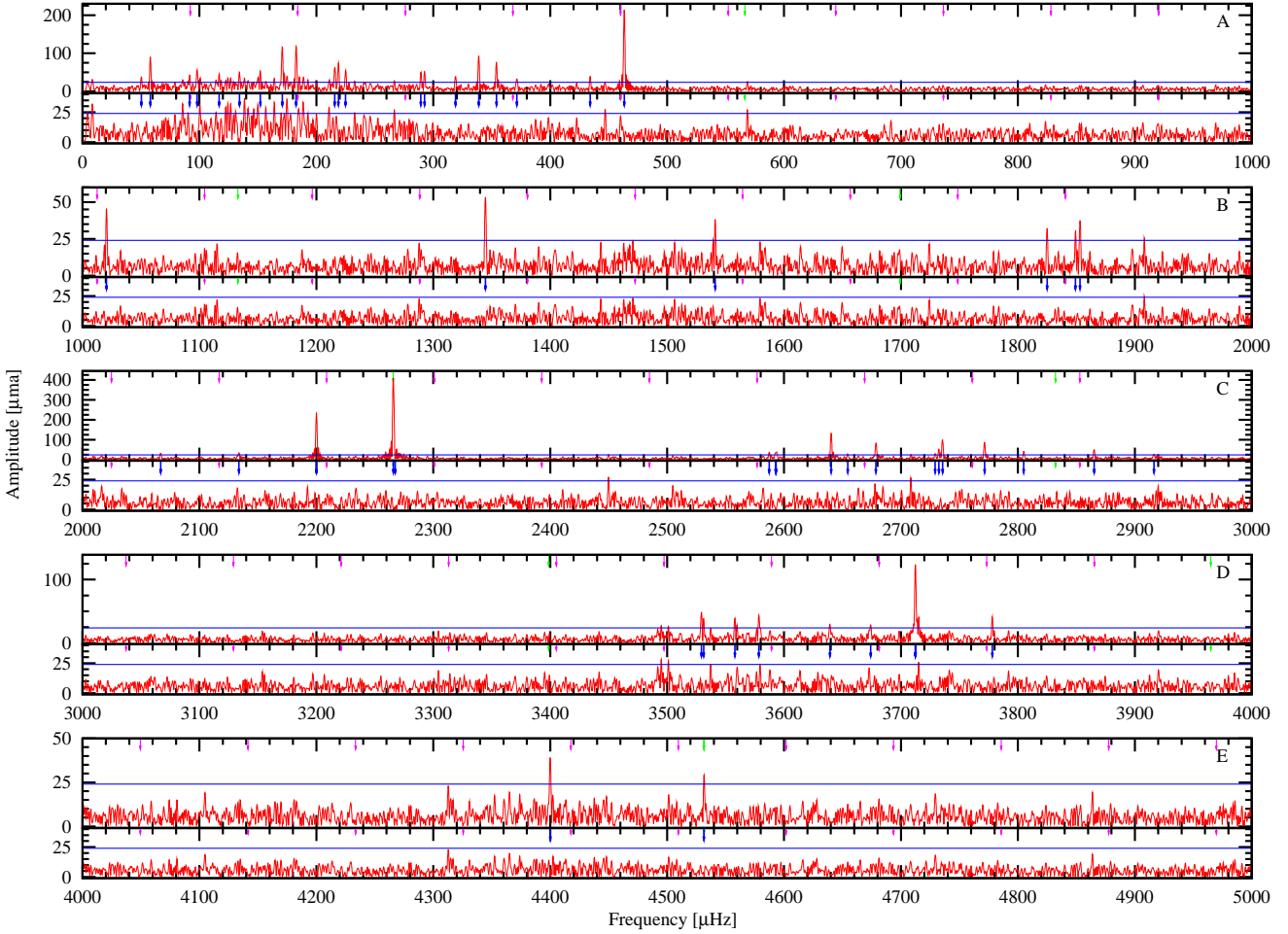}
   \caption{The FT of the data after detrending the orbital and long
            term variations (upper part of each panel). The panels are
            labeled {\sc a,\,b,\,c,\,d,\,e} and these also define the group
            designation used in Table~\ref{tbl:freqs}.  The magenta
            arrows indicate the orbital frequency of \ueq{\forb}{92}{\uHz}~and
            all its harmonics, which have been completely removed
            by subtracting the data folded \forb.
            The green arrows indicate the \kep\ long-cadence cycle, \flc~and
            its harmonics, at which possible artefacts are known to be found.
            The lower part of each panel shows the FT of the data
            after prewhitening 55 frequencies, as marked with blue
            arrows. The blue line indicates four times the mean noise
            level at 24\,\uma. 
           }
      \label{fig:ft}
\end{figure*}

In order to produce a useful Fourier Transform (FT) that shows
the spectrum of low-level pulsations in \target\ among the extremely
dominant orbital effects (Fig.~\ref{fig:lc}, top panel) we first
attempted to clean out a model light curve of the system, but
we abandoned this approach since the FT of the residuals
contain significant peaks at {\em every} orbital harmonic.
The model light curve is unable to produce a satisfactory
fit to the complicated irradiation effect, at the exceptional
precision of the \kep\ photometry.
Our model light curve does not account for radiative transfer through
the heated face of the M-dwarf, which may account for some of the
discrepancies between the model and data.
Until such issues are resolved the precise parameters of our model are
subject to systematic uncertainties that could well be in excess of the
statistical errors.
Even if we managed to model all orbital effects in the light curve to the
required precision, any pulsation peaks found in the residuals on an orbital
harmonic frequency would still be suspicious. So instead we proceeded by
folding the \kep\ light curve on $P$, and then using the result to clean
out all orbital effects from the light curve.  Note
that any stable pulsations that coincide with these periods within the
resolution are also very effectively removed.

\subsection{Frequency determination}

After removing the orbital effects from the light curve, a spectacularly
rich pulsation spectrum is revealed (upper half of the five panels in
Fig.~\ref{fig:ft}). We have identified 55 frequencies between $\sim$50 and
4500\,\uHz\ that have amplitudes well above four times the mean level
of 6\,\uma\footnote{One micro-modulation amplitude (\uma) is equivalent
to a semi-amplitude of one part per million of the mean light level.}
computed from the whole FT between 500\,\uHz\ and the Nyquist frequency,
\fnyq\,=\,8496\,\uHz. These frequencies are listed in Table~\ref{tbl:freqs}.
Several more frequencies remain above the 24\,\uma\ limit,
after prewhitening the light curve with these 55 frequencies
(lower half of the panels in Fig.~\ref{fig:ft}), especially in the
low frequency region below 500\,\uHz. However, we chose to constrain
ourselves to the most clearly resolved peaks in this first analysis,
as much more \kep\ data is underway, and will provide a tenfold
increase in resolution and a threefold drop in the noise level after
only the first three-month cycle of observations. In total we hope to
obtain close to five years of almost continuous space based observations
on this object, so we will limit ourselves to discussing only the
most obvious features here.

The FT shows several distinct groups of frequencies, and we number
the frequencies in each group separately as {\sc a,\,b,\,c,\,d,\,e}, as
indicated in Fig.~\ref{fig:ft}. The frequencies in group {\sc a} are
typical for the V1093\,Her stars, and we identify and remove 20 peaks,
with the lowest at 6\,\sig. There are clearly many more frequencies
than this, in particular between 100 and 200\,\uHz, but the resolution
is not good enough to justify pushing the limit further.
There are no significant
frequencies between 500 and 1000\,\uHz, except one at 570\,\uHz\ which
is just above the detection limit. Note also that this frequency
is close to the long cadence readout cycle at \flc\,=\,566.4\,\uHz.
In group {\sc b} we find only six rather low-level peaks.
Group {\sc c} contains the strongest peak in the FT at
$f_{C4}$\,=\,2265.8\,\uHz\ with an amplitude of 423\,\uma. This frequency
is almost exactly at 4\flc\,=\,2265.4\,\uHz, but since we do not
observe such strong artefacts in any other stars we are inclined
to believe that this is a real pulsation mode. In all
other short cadence \kep\ datasets we have worked with, the strongest
artefact peak is found at 8 or 9\flc. In this dataset 9\flc\ is not
present at all, and 8\flc\,=\,4531\,\uHz\ is barely above the detection
limit, while $f_{C4}$ is 70-\sig! The second strongest peak in the FT is also
found right next to it, $f_{C3}$\,=\,2200.1\,\uHz\
with an amplitude of 233\,\uma.
Another rich set of peaks is found around 2700\,\uHz, so that group {\sc c}
contains 18 peaks altogether.
Group {\sc d} contains 9 clear peaks between 3500 and 3800\,\uHz, and
group {\sc e} contains only two significant peaks located at 2$f_{C3}$ and
2$f_{C4}$\,=\,8\flc.
The ratio of some of the high-amplitude peaks in the short-period range,
such as $f_{D7}/f_{C4}$\,=\,1.638 and $f_{D9}/f_{C3}$\,=\,1.717, which are
both close to the theoretical ratio of the radial fundamental and the
second overtone (1.667), and a stellar model for the \teff\ and \logg\ 
of the primary does display pulsation periods for radial modes close
to these frequencies.

The frequency analysis and error estimations on the parameters are done
as outlined in \citet{degroote2009b}. Because some of the frequencies
were closer together than the Rayleigh $T^{-1}$ limit, and thus unresolved,
we have performed additional analyses, where after each prewhitening stage,
some of the frequencies, amplitudes and phases were refitted with a
Levenberg--Marquardt nonlinear fitting routine.
To investigate the stability of the fit and the differences from the
linear analysis, we first only updated the amplitudes, phases and
frequencies belonging the last two identified peaks, then updating
the parameters belonging to the last 20 peaks, and finally using all
identified frequencies. The differences between the amplitudes and
frequencies were all within the derived error estimates, except for the
frequencies around 3712 and 2593\,\uHz. The
nonlinear fitting algorithm did not produce consistent results at these
frequencies, but future observations will allow us to resolve these structures.
Below 200\,\uHz, a dense forest of barely significant peaks is visible.
Although the different methods give consistent results,
more \kep\ photometry will be necessary to fully resolve them.

\subsection{Spectroscopy}

Spectroscopic observations to determine radial velocities for \target\ were
undertaken primarily with the B\&C spectrograph on the 2.3-m
{\em Bok Telescope} on Kitt Peak. A few spectra were also obtained
using the MMT Blue spectrograph. In both cases, an 832/mm grating
was employed in second order, and particular care was taken to maintain
precise centering of the star on the slit throughout the exposures.
The resulting spectra have a resolution of $R$\,$\sim$\,2150 over the range
3675\,--\,4520\,\AA~(Bok), or $R$\,$\sim$\,4200, 4000\,--\,4950\,\AA~(MMT).
Radial velocities
were derived by cross-correlating the individual continuum-removed
spectra against a super-template using the {\sc iraf} 
task {\tt FXCOR}. 
The super-template for the lower resolution spectra was obtained by
shifting the 66 individual spectra to the same velocity prior to median
filtering into a single spectrum; the velocity zero point of the resulting
template was thus undetermined.  The cross-correlation template for
the 6 MMT spectra was constructed from 19 spectra previously obtained
with the identical spectroscopic setup for other hot subdwarfs of
known radial velocities, whose spectral abundance patterns closely
match that of 2M1938+4603.

A single sinusoidal fit was performed to all 72 velocities as a
function of orbital phase, weighted by the velocity errors and also
including an additional term for the zero-point offset of the Bok
velocities relative to the MMT velocities (Fig.~\ref{fig:rv}). 
The orbital period was fixed at the value of 0.12576530\,d derived from the
eclipse timings.  The derived RV semi-amplitude is
\eueq{K_1}{65.7}{0.6}{\kms},
with a systemic velocity
\eueq{\gamma}{20.1}{0.3}{\kms}.

\begin{figure}
\includegraphics[width=\hsize]{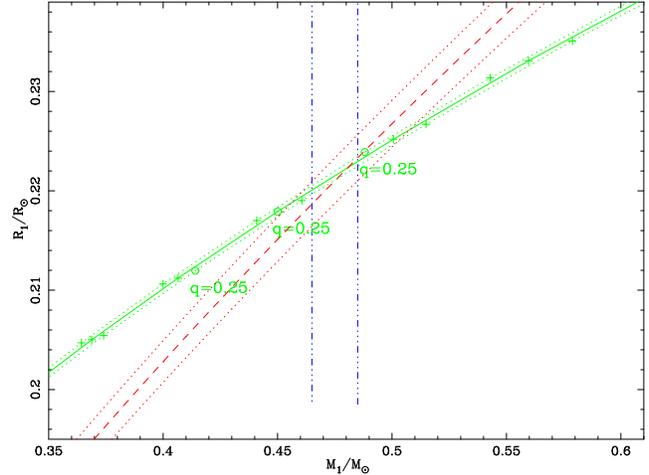}
\caption{
Mass--radius diagram for \target\ showing the regions permitted by
the orbit solution (green curves) and by the spectroscopic
gravity (red curves).
The dotted, green curves show the error ranges that corresponds to
a 3-$\sigma$ error in $K_1$ for the orbit solution. The dotted, red curves
corresponds to the one \sig~formal error on \logg\ for the
gravity track. The green curves are labeled with the value for
$q$\,=\,$M_1$/$M_2$ that corresponds to each point on the curve,
with + marking ticks of 0.01, increasing to the left.
Note that changing $K_1$ by 3$\sigma$ shifts the curve very
little in the $M$/$R$ plane, but the $q$-value changes considerably. 
Vertical lines show the mass range typical for canonical EHB stars formed
through the CE channel.
}
\label{fig:mr1}
\end{figure}

\section{System parameters}

The effective temperature, surface gravity, and helium abundance
of \target\ were determined in the context of Paper\,{\sc i} to be
\eueq{\teff}{29\,564}{106}{K},
\eueq{\logg}{5.425}{\,0.009}{dex}, and
\eueq{\lheh}{--2.36}{0.06}{dex},
using metal blanketed LTE models. This temperature is close to the
boundary region between the $p$- and $g$-mode pulsators, where
hybrid DW\,Lyn type pulsators have been found.

We modelled the light curve using grid elements covering each star,
accounting for tidal (ellipsoidal) distortion of each star and gravity
and limb-darkening of the sdB as in \citet{bloemen10}.
Irradiation of the M-dwarf was accounted for by summing the M-dwarf flux
with the incident flux from the sdB point by point.
Surface brightnesses were computed assuming black-body spectra at a
single wavelength of 600\,nm.
The best fitting model is found to have the following parameters:
inclination angle, $i$\,=\,69.45(2)$^\circ$, relative
radii, $r_1$\,=\,0.250(1), $r_2$\,=\,0.177(1) in units of
the orbital separation. The errors in parentheses are formal fitting
errors that may underestimate the true errors perhaps as much as a
factor 10, considering the discrepancies in the light curve fitting.
The high precision on these parameters allows us to use
the mass-radius relationships as derived from the orbital parameters
and from the surface gravity to constrain the mass and radius of the primary,
as shown in Fig.~\ref{fig:mr1}.
We clearly see that the permitted $M$/$R$ given by the photometric orbital
parameters ($P$, $i$, $r_1$ and $r_2$) and the spectroscopic $K_1$
crosses the $M$/$R$ given by the spectroscopic surface gravity at
precisely the expected primary mass for a post-CE sdB star.
The adjacent dotted lines indicate the errors (on the RV for the
$q$-track, and on \logg\ for the $g$-track since the errors on $P$, $i$
and the relative radii are too small to matter).
A mass for the primary of \eueq{M_1}{0.48}{0.03}{\msol} can be deduced
from the diagram.
Using $K_1$ and $P$ we know the mass function $f(M)$\,=\,0.003695.
With 0.48\,\msol\ for the primary, and solving for the secondary,
we get \eueq{M_2}{0.12}{0.01}{\msol}.

\section{Discussion and Outlook}

We have presented the discovery of a new eclipsing sdB with an M-dwarf
companion, and demonstrated that the primary is an unusual pulsator
with frequencies spanning the range from 50 to 4400\,\uHz. 
The unprecedented precision of the \kep\ observations makes a direct
comparison to other known eclipsing sdB+dM systems rather difficult,
since such low-level pulsations as we find in \target\ would be
undetectable from the ground. Nevertheless, it is an amazing stroke
of luck to find an eclipsing sdB+dM system within the narrow
confines of the \kep\ field, especially one as bright as 12th magnitude,
considering the fact that only seven
such systems are known among the thousands of subdwarfs that have been
surveyed for variability to date.

In future papers we will produce a more realistic model that can
handle the irradiation effect of the M-dwarf to improve the binary light curve
model, and produce asteroseismic models that can reproduce the
frequencies observed in this star. When longer time-series of \kep\
photometry become available, we will be able to fully resolve the
densely packed amplitude spectrum that we see in the $g$-mode region
between 100 and 200\,\uHz. This will enable us to produce an
asteroseismic model reproducing all the frequencies in this complex
rotating pulsator, as was done for NY\,Vir by \citet{charpinet08}.
Since \target\ has $g$-modes in addition to $p$-modes we should also be
able to constrain the deeper structure of the sdB core,
and hopefully also establish the mass of its
progenitor, before the envelope was ejected by the companion.
With five years of precise eclipse timings we may
detect planetary companions from variations in the O--C diagram,
if they orbit this compact binary with periods shorter than the \kepmi,
and the low-amplitude pulsations are found to be stable on such time-scales.

\section*{Acknowledgments}

The authors gratefully acknowledge the \kep\ team and everybody who
has contributed to making this mission possible.
Funding for the \kepmi\ is provided by NASA's Science Mission Directorate.

Part of the spectroscopic observations reported here were obtained at the
MMT Observatory, a joint facility of the University of Arizona and the
Smithsonian Institution.

The research leading to these results has received funding from the European
Research Council under the European Community's Seventh Framework Programme
(FP7/2007--2013)/ERC grant agreement N$^{\underline{\mathrm o}}$\,227224
({\sc prosperity}), as well as from the Research Council of K.U.Leuven grant
agreement GOA/2008/04.

\bibliographystyle{mn2e}
\bibliography{sdbrefs}

\begin{thebibliography}{}

\bibitem[\protect\citeauthoryear{{Bloemen}, {Marsh}, {\O stensen}, {Charpinet},
  {Degroote}, {Kawaler}, {Aerts}, {Fontaine}, {Green}, {Telting}, {Brassard}
  et~al.,}{{Bloemen} et~al.}{2010}]{bloemen10}
{Bloemen} S. et~al., 2010, \mnras, submitted, ...

\bibitem[\protect\citeauthoryear{{Borucki}, {Koch}, {Basri}, {Batalha},
  {Brown}, {Caldwell}, {Caldwell}, {Christensen-Dalsgaard}, {Cochran},
  {DeVore}, {Dunham}, {Dupree}, {Gautier}, {Geary}, {Gilliland}
  et~al.,}{{Borucki} et~al.}{2010}]{borucki10}
{Borucki} W.~J. et~al., 2010, Science, 327, 977

\bibitem[\protect\citeauthoryear{{Charpinet}, {Fontaine}, {Brassard}, {Chayer},
  {Rogers}, {Iglesias} \& {Dorman}}{{Charpinet} et~al.}{1997}]{charpinet97}
{Charpinet} S., {Fontaine} G., {Brassard} P., {Chayer} P., {Rogers} F.~J.,
  {Iglesias} C.~A., {Dorman} B. 1997, \apjl, 483, L123

\bibitem[\protect\citeauthoryear{{Charpinet}, {van Grootel}, {Reese},
  {Fontaine}, {Green}, {Brassard} \& {Chayer}}{{Charpinet}
  et~al.}{2008}]{charpinet08}
{Charpinet} S., {van Grootel} V., {Reese} D., {Fontaine} G., {Green} E.~M.,
  {Brassard} P., {Chayer} P. 2008, \aap, 489, 377

\bibitem[\protect\citeauthoryear{{Degroote}, {Aerts}, {Ollivier}, {Miglio},
  {Debosscher}, {Cuypers}, {Briquet}, {Montalb{\'a}n}, {Thoul}, {Noels}, {De
  Cat}, {Balaguer-N{\'u}{\~n}ez}, {Maceroni}, {Ribas}, {Auvergne}, {Baglin}
  et~al.,}{{Degroote} et~al.}{2009}]{degroote2009b}
{Degroote} P. et~al., 2009, \aap, 506, 471

\bibitem[\protect\citeauthoryear{{Fontaine}, {Brassard}, {Charpinet}, {Green},
  {Chayer}, {Bill{\`e}res} \& {Randall}}{{Fontaine} et~al.}{2003}]{fontaine03}
{Fontaine} G., {Brassard} P., {Charpinet} S., {Green} E.~M., {Chayer} P.,
  {Bill{\`e}res} M., {Randall} S.~K. 2003, \apj, 597, 518

\bibitem[\protect\citeauthoryear{{For}, {Green}, {Fontaine}, {Drechsel},
  {Shaw}, {Dittmann}, {Fay}, {Francoeur}, {Laird}, {Moriyama}, {Morris},
  {Rodr{\'{\i}}guez-L{\'o}pez}, {Sierchio}, {Story}, {Strom}, {Wang}
  et~al.,}{{For} et~al.}{2010}]{for10}
{For} B.-Q. et~al., 2010, \apj, 708, 253

\bibitem[\protect\citeauthoryear{{Gilliland}, {Brown}, {Christensen-Dalsgaard},
  {Kjeldsen}, {Aerts}, {Appourchaux}, {Basu}, {Bedding}, {Chaplin}, {Cunha},
  {De Cat}, {De Ridder}, {Guzik}, {Handler} et~al.,}{{Gilliland}
  et~al.}{2010}]{gilliland10a}
{Gilliland} R.~L. et~al., 2010, \pasp, 122, 131

\bibitem[\protect\citeauthoryear{{Green}, {Fontaine}, {Reed}, {Callerame},
  {Seitenzahl}, {White}, {Hyde}, {{\O}stensen}, {Cordes}, {Brassard}, {Falter},
  {Jeffery}, {Dreizler}, {Schuh}, {Giovanni}, {Edelmann}, {Rigby} \&
  {Bronowska}}{{Green} et~al.}{2003}]{green03}
{Green} E.~M. et~al., 2003, \apjl, 583, L31

\bibitem[\protect\citeauthoryear{{Han}, {Podsiadlowski}, {Maxted}, {Marsh} \&
  {Ivanova}}{{Han} et~al.}{2002}]{han02}
{Han} Z., {Podsiadlowski} P., {Maxted} P.~F.~L., {Marsh} T.~R., {Ivanova} N.
  2002, \mnras, 336, 449

\bibitem[\protect\citeauthoryear{{Heber}}{{Heber}}{1986}]{heber86}
{Heber} U. 1986, \aap, 155, 33

\bibitem[\protect\citeauthoryear{{Heber}}{{Heber}}{2009}]{heber09}
{Heber} U. 2009, \araa, 47, 211

\bibitem[\protect\citeauthoryear{{Kawaler}, {Reed}, {Quint}, {\O stensen},
  {Silvotti}, {Baran}, {Charpinet}, {Bloemen}, {Kurtz}, {Telting}, {Handler}
  et~al.,}{{Kawaler} et~al.}{2010a}]{kawaler10a}
{Kawaler} S.~D. et~al., 2010a, \mnras, submitted, ...

\bibitem[\protect\citeauthoryear{{Kawaler}, {Reed}, {\O stensen}, {Bloemen},
  {Kurtz}, {Quint}, {Silvotti}, {Baran}, {Green}, {Charpinet}, {Telting},
  {Aerts}, {Handler} et~al.,}{{Kawaler} et~al.}{2010b}]{kawaler10b}
{Kawaler} S.~D. et~al., 2010b, \mnras, submitted, ...

\bibitem[\protect\citeauthoryear{{Kilkenny}, {Koen}, {O'Donoghue} \&
  {Stobie}}{{Kilkenny} et~al.}{1997}]{kilkenny97}
{Kilkenny} D., {Koen} C., {O'Donoghue} D., {Stobie} R.~S. 1997, \mnras, 285,
  640

\bibitem[\protect\citeauthoryear{{Kilkenny}, {O'Donoghue}, {Koen}, {Lynas-Gray}
  \& {van Wyk}}{{Kilkenny} et~al.}{1998}]{kilkenny98}
{Kilkenny} D., {O'Donoghue} D., {Koen} C., {Lynas-Gray} A.~E., {van Wyk} F.
  1998, \mnras, 296, 329

\bibitem[\protect\citeauthoryear{{Kilkenny}, {van Wyk} \& {Marang}}{{Kilkenny}
  et~al.}{2003}]{kilkenny03_HWVir}
{Kilkenny} D., {van Wyk} F., {Marang} F. 2003, The Observatory, 123, 31

\bibitem[\protect\citeauthoryear{{Lee}, {Kim}, {Kim}, {Koch}, {Lee}, {Kim} \&
  {Park}}{{Lee} et~al.}{2009}]{lee09}
{Lee} J.~W., {Kim} S.-L., {Kim} C.-H., {Koch} R.~H., {Lee} C.-U., {Kim} H.-I.,
  {Park} J.-H. 2009, \aj, 137, 3181

\bibitem[\protect\citeauthoryear{{\O stensen}}{{\O stensen}}{2009}]{ostensen09}
{\O stensen} R. 2009, Communications in Asteroseismology, 159, 75

\bibitem[\protect\citeauthoryear{{\O stensen}, {Oreiro}, {Solheim}, {Heber},
  {Silvotti}, {Gonz{\'a}lez-P{\'e}rez}, {Ulla}, {P{\'e}rez Hern{\'a}ndez},
  {Rodr{\'{\i}}guez-L{\'o}pez} \& {Telting}}{{\O stensen}
  et~al.}{2010a}]{sdbnot}
{\O stensen} R.~H. et~al., 2010a, \aap, 513, A6

\bibitem[\protect\citeauthoryear{{\O stensen}, {Silvotti}, {Charpinet},
  {Oreiro}, {Handler}, {Green}, {Bloemen}, {Heber}, {G\"ansicke}, {Marsh}
  et~al.,}{{\O stensen} et~al.}{2010b}]{kepI}
{\O stensen} R.~H. et~al., 2010b, \mnras, submitted, ...

\bibitem[\protect\citeauthoryear{{Reed}, {Kawaler}, {\O stensen}, {Bloemen},
  {Baran}, {Silvotti}, {Charpinet}, {Quint}, {Handler}, {Gilliland}
  et~al.,}{{Reed} et~al.}{2010}]{reed10a}
{Reed} M. et~al., 2010, \mnras, submitted, ...

\bibitem[\protect\citeauthoryear{{Schuh}, {Huber}, {Dreizler}, {Heber},
  {O'Toole}, {Green} \& {Fontaine}}{{Schuh} et~al.}{2006}]{schuh06}
{Schuh} S., {Huber} J., {Dreizler} S., {Heber} U., {O'Toole} S.~J., {Green}
  E.~M., {Fontaine} G. 2006, \aap, 445, L31

\bibitem[\protect\citeauthoryear{{Skrutskie}, {Cutri}, {Stiening}, {Weinberg},
  {Schneider}, {Carpenter}, {Beichman}, {Capps}, {Chester}, {Elias}, {Huchra},
  {Liebert}, {Lonsdale}, {Monet} et~al.,}{{Skrutskie} et~al.}{2006}]{twomass}
{Skrutskie} M.~F. et~al., 2006, \aj, 131, 1163

\bibitem[\protect\citeauthoryear{{Van Cleve}}{{Van Cleve}}{2009}]{kepler_dr02}
{Van Cleve} J.~E. 2009, {Kepler Data Release Notes 2},
  http://archive.stsci.edu/kepler/release\_notes/release\_notes2/\\Data\_Relea%
se\_02\_Notes\_2009102213.pdf

\bibitem[\protect\citeauthoryear{{Van Grootel}, {Charpinet}, {Fontaine},
  {Brassard}, {Green}, {Randall}, {Silvotti}, {\O stensen}, {Gilliland}
  et~al.,}{{Van Grootel} et~al.}{2010}]{VanGrootel10}
{Van Grootel} V. et~al., 2010, ApJL, submitted, ...

\end{thebibliography}

\label{lastpage}

\clearpage
\appendix
\section{Ephemeris}

\begin{table}
\centering
\caption{Times of primary minima from ground based observations, and
O--C values for $T_0$\,=\,54640.864162, $P$\,=\,0.125765300.}
\label{tbl:ephem1}
\small
\begin{tabular}{rrrrc}\hline\hline
   \multicolumn{1}{c}{Epoch} &
   \multicolumn{1}{c}{BJD} &
   \multicolumn{1}{c}{$\sigma_\mathrm{BJD}$} &
   \multicolumn{1}{c}{O--C} \\ \hline
   0 & 54640.86420 &  0.00004 &  0.00004 \\
 595 & 54715.69440 &  0.00004 & -0.00012 \\
 714 & 54730.66060 &  0.00004 &  0.00001 \\
 929 & 54757.70010 &  0.00004 & -0.00003 \\
2569 & 54963.95530 &  0.00004 &  0.00008 \\
2577 & 54964.96140 &  0.00004 &  0.00006 \\
2712 & 54981.93960 &  0.00004 & -0.00006 \\
2720 & 54982.94560 &  0.00004 & -0.00018 \\
2767 & 54988.85690 &  0.00004 &  0.00015 \\
2775 & 54989.86305 &  0.00004 &  0.00018 \\
3434 & 55072.74210 &  0.00004 & -0.00010 \\
3696 & 55105.69270 &  0.00004 & -0.00001 \\
5527 & 55335.96895 &  0.00004 & -0.00003 \\
\hline
\end{tabular}\end{table}

\begin{table}
\centering
\caption{Times of primary minima from \kep\ observations, and
O--C values for $T_0$\,=\,54640.864162, $P$\,=\,0.125765300.}
\label{tbl:ephem2}
\small
\begin{tabular}{rrrrc}\hline\hline
   \multicolumn{1}{c}{Epoch} &
   \multicolumn{1}{c}{BJD} &
   \multicolumn{1}{c}{$\sigma_\mathrm{BJD}$} &
   \multicolumn{1}{c}{O--C} \\ \hline
2487 & 54953.64250 &  0.00001 &  0.00004 \\
2488 & 54953.76824 &  0.00001 &  0.00001 \\
2489 & 54953.89401 &  0.00001 &  0.00002 \\
2490 & 54954.01972 &  0.00001 & -0.00004 \\
2491 & 54954.14554 &  0.00001 &  0.00001 \\
2492 & 54954.27131 &  0.00001 &  0.00002 \\
2493 & 54954.39705 &  0.00001 & -0.00001 \\
2494 & 54954.52283 &  0.00001 &  0.00001 \\
2495 & 54954.64861 &  0.00001 &  0.00002 \\
2496 & 54954.77433 &  0.00001 & -0.00002 \\
2497 & 54954.90010 &  0.00001 & -0.00002 \\
2498 & 54955.02588 &  0.00001 & -0.00000 \\
2499 & 54955.15162 &  0.00001 & -0.00003 \\
2500 & 54955.27741 &  0.00001 & -0.00000 \\
2501 & 54955.40317 &  0.00001 & -0.00000 \\
2502 & 54955.52895 &  0.00001 &  0.00000 \\
2503 & 54955.65467 &  0.00001 & -0.00004 \\
2504 & 54955.78047 &  0.00001 & -0.00001 \\
2505 & 54955.90625 &  0.00001 &  0.00001 \\
2506 & 54956.03200 &  0.00001 & -0.00000 \\
2507 & 54956.15777 &  0.00001 & -0.00000 \\
2508 & 54956.28355 &  0.00001 &  0.00001 \\
2509 & 54956.40926 &  0.00001 & -0.00004 \\
2510 & 54956.53506 &  0.00001 & -0.00001 \\
2511 & 54956.66083 &  0.00001 & -0.00000 \\
2512 & 54956.78659 &  0.00001 & -0.00001 \\
2513 & 54956.91237 &  0.00001 &  0.00001 \\
2514 & 54957.03814 &  0.00001 &  0.00001 \\
2515 & 54957.16392 &  0.00001 &  0.00003 \\
2516 & 54957.28972 &  0.00001 &  0.00006 \\
2517 & 54957.41545 &  0.00001 &  0.00003 \\
2518 & 54957.54121 &  0.00001 &  0.00002 \\
2519 & 54957.66696 &  0.00001 &  0.00001 \\
2520 & 54957.79274 &  0.00001 &  0.00002 \\
2521 & 54957.91850 &  0.00001 &  0.00001 \\
2522 & 54958.04428 &  0.00001 &  0.00003 \\
2523 & 54958.17004 &  0.00001 &  0.00003 \\
2524 & 54958.29579 &  0.00001 &  0.00001 \\
2525 & 54958.42152 &  0.00001 & -0.00002 \\
2526 & 54958.54734 &  0.00001 &  0.00003 \\
2527 & 54958.67311 &  0.00001 &  0.00003 \\
2528 & 54958.79883 &  0.00001 & -0.00001 \\
2529 & 54958.92465 &  0.00001 &  0.00004 \\
2530 & 54959.05035 &  0.00001 & -0.00002 \\
2531 & 54959.17612 &  0.00001 & -0.00002 \\
2532 & 54959.30187 &  0.00001 & -0.00003 \\
2533 & 54959.42766 &  0.00001 & -0.00001 \\
2534 & 54959.55344 &  0.00001 &  0.00001 \\
2535 & 54959.67918 &  0.00001 & -0.00001 \\
2536 & 54959.80498 &  0.00001 &  0.00002 \\
2537 & 54959.93071 &  0.00001 & -0.00001 \\
2538 & 54960.05651 &  0.00001 &  0.00001 \\
2539 & 54960.18225 &  0.00001 & -0.00001 \\
2540 & 54960.30802 &  0.00001 &  0.00000 \\
2541 & 54960.43378 &  0.00001 & -0.00001 \\
2542 & 54960.55957 &  0.00001 &  0.00002 \\
2543 & 54960.68531 &  0.00001 & -0.00001 \\
2544 & 54960.81113 &  0.00001 &  0.00004 \\
2545 & 54960.93685 &  0.00001 & -0.00000 \\
2546 & 54961.06262 &  0.00001 &  0.00001 \\
\hline
\end{tabular}\end{table}
\begin{table}
\centering
\caption{Times of primary minima from \kep\ observations, cont. }
\small
\begin{tabular}{rrrrc}\hline\hline
   \multicolumn{1}{c}{Epoch} &
   \multicolumn{1}{c}{BJD} &
   \multicolumn{1}{c}{$\sigma_\mathrm{BJD}$} &
   \multicolumn{1}{c}{O--C} \\ \hline
2547 & 54961.18842 &  0.00001 &  0.00004 \\
2548 & 54961.31416 &  0.00001 &  0.00001 \\
2549 & 54961.43995 &  0.00001 &  0.00004 \\
2550 & 54961.56569 &  0.00001 &  0.00001 \\
2551 & 54961.69145 &  0.00001 &  0.00001 \\
2552 & 54961.81719 &  0.00001 & -0.00002 \\
2553 & 54961.94298 &  0.00001 &  0.00001 \\
2554 & 54962.06873 &  0.00001 & -0.00001 \\
2555 & 54962.19452 &  0.00001 &  0.00002 \\
2556 & 54962.32026 &  0.00001 & -0.00001 \\
2557 & 54962.44603 &  0.00001 & -0.00000 \\
2558 & 54962.57180 &  0.00001 &  0.00000 \\
2559 & 54962.69755 &  0.00001 & -0.00001 \\
2560 & 54962.82334 &  0.00001 &  0.00001 \\
2561 & 54962.94909 &  0.00001 & -0.00001 \\
2562 & 54963.07486 &  0.00001 & -0.00000 \\
2563 & 54963.20060 &  0.00001 & -0.00002 \\
\hline
\end{tabular}\end{table}

Online material.


\clearpage
\section{RV measurements}

\begin{figure}
\centering
\includegraphics[height=\hsize,clip,angle=-90]{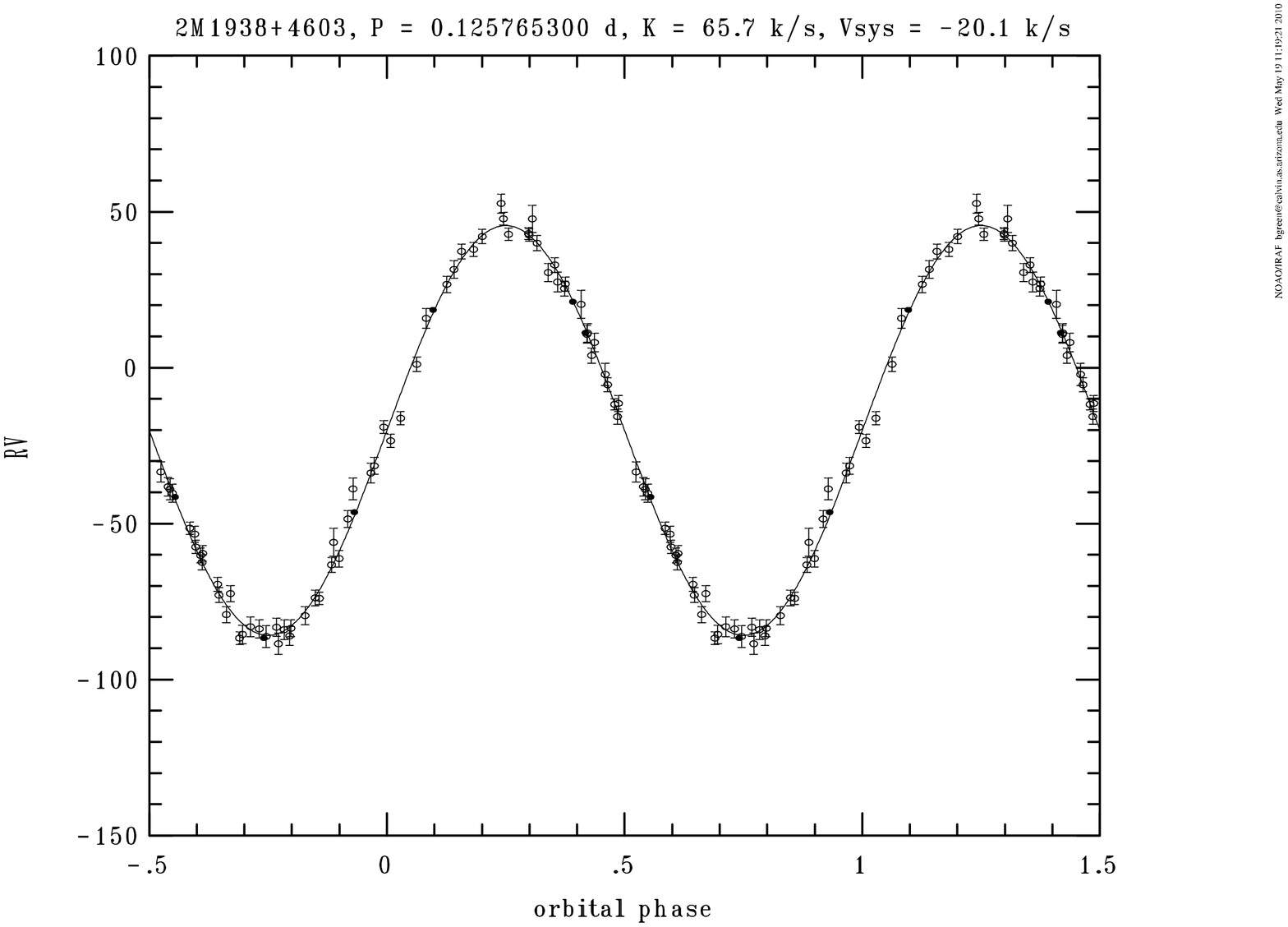}
\caption{The spectroscopic RVs. Open points are from the {\em
Bok Telescope}, filled dots from the MMT.  }
      \label{fig:rv}
\end{figure}

\section{Light curve model}

\begin{figure}
\includegraphics[height=\hsize,angle=-90]{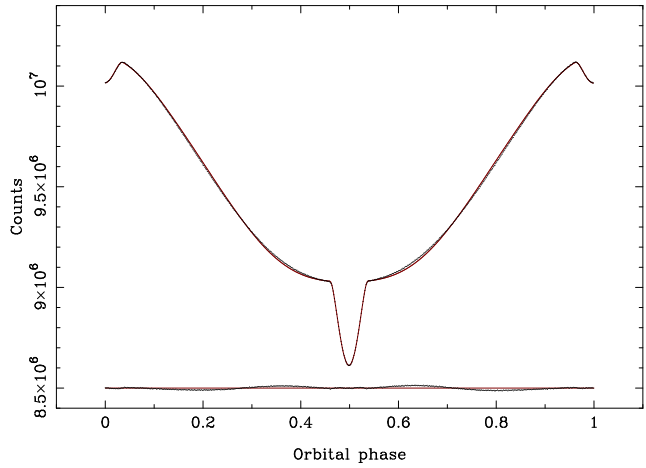}
\caption{
Best fit model light curve and the folded \kep\ light curve (top),
and residuals (bottom).
}
\label{fig:fit1}
\end{figure}

\clearpage

\section{Frequency list}

\begin{table}[h!]
\centering
\caption{Clearly resolved frequencies, periods, and amplitudes for \target.}
\label{tbl:freqs}\small
\begin{tabular}{lrrrc}\hline\hline
   \multicolumn{1}{c}{ID} &
   \multicolumn{1}{c}{Frequency} &
   \multicolumn{1}{c}{Period} &
   \multicolumn{1}{c}{Amplitude} \\
   &
   \multicolumn{1}{c}{[\uHz]} &
   \multicolumn{1}{c}{[s]} &
   \multicolumn{1}{c}{[\uma]} \\ \hline
$f_{A1}$ &   50.289 (137) & 19884.934 (54.225) &   36 (7) \\
$f_{A2}$ &   57.959 (056) & 17253.692 (16.761) &   89 (7) \\
$f_{A3}$ &   91.614 (123) & 10915.351 (14.709) &   40 (7) \\
$f_{A4}$ &   97.954 (089) & 10208.828 (09.325) &   56 (7) \\
$f_{A5}$ &  116.859 (108) &  8557.319 (07.969) &   46 (7) \\
$f_{A6}$ &  134.181 (099) &  7452.603 (05.500) &   50 (7) \\
$f_{A7}$ &  152.043 (090) &  6577.086 (03.923) &   55 (7) \\
$f_{A8}$ &  170.824 (043) &  5853.966 (01.474) &  116 (7) \\
$f_{A9}$ &  182.673 (042) &  5474.256 (01.260) &  119 (7) \\
$f_{A10}$ &  215.762 (085) &  4634.736 (01.834) &   59 (7) \\
$f_{A11}$ &  218.939 (070) &  4567.474 (01.472) &   72 (7) \\
$f_{A12}$ &  224.989 (096) &  4444.656 (01.898) &   52 (7) \\
$f_{A13}$ &  289.349 (113) &  3456.036 (01.352) &   45 (7) \\
$f_{A14}$ &  292.537 (102) &  3418.370 (01.201) &   49 (7) \\
$f_{A15}$ &  319.054 (121) &  3134.262 (01.191) &   41 (7) \\
$f_{A16}$ &  338.857 (054) &  2951.098 (00.472) &   92 (7) \\
$f_{A17}$ &  353.995 (066) &  2824.900 (00.533) &   75 (7) \\
$f_{A18}$ &  371.400 (153) &  2692.512 (01.113) &   32 (7) \\
$f_{A19}$ &  434.133 (129) &  2303.443 (00.688) &   38 (7) \\
$f_{A20}$ &  463.297 (023) &  2158.441 (00.108) &  214 (7) \\
$f_{B1}$ & 1020.507 (110) &   979.906 (00.106) &   45 (7) \\
$f_{B2}$ & 1344.545 (093) &   743.746 (00.052) &   53 (7) \\
$f_{B3}$ & 1541.048 (131) &   648.909 (00.055) &   38 (7) \\
$f_{B4}$ & 1824.991 (158) &   547.948 (00.048) &   31 (7) \\
$f_{B5}$ & 1849.248 (164) &   540.760 (00.048) &   31 (7) \\
$f_{B6}$ & 1853.196 (134) &   539.608 (00.039) &   37 (7) \\
$f_{C1}$ & 2066.784 (156) &   483.843 (00.037) &   32 (7) \\
$f_{C2}$ & 2133.776 (155) &   468.653 (00.034) &   32 (7) \\
$f_{C3}$ & 2200.061 (021) &   454.533 (00.004) &  233 (7) \\
$f_{C4}$ & 2265.789 (016) &   441.347 (00.003) &  423 (7) \\
$f_{C5}$ & 2267.414 (110) &   441.031 (00.022) &   63 (7) \\
$f_{C6}$ & 2587.292 (150) &   386.504 (00.022) &   34 (7) \\
$f_{C7}$$^\dag$ & 2592.837 (134) &   385.678 (00.020) &  40 (8) \\
$f_{C8}$$^\dag$ & 2593.186 (134) &   385.626 (00.020) &  -- (--) \\
$f_{C9}$ & 2640.303 (037) &   378.744 (00.005) &  133 (7) \\
$f_{C10}$ & 2654.473 (172) &   376.723 (00.024) &   29 (7) \\
$f_{C11}$ & 2678.609 (059) &   373.328 (00.008) &   84 (7) \\
$f_{C12}$ & 2729.208 (142) &   366.407 (00.019) &   35 (7) \\
$f_{C13}$ & 2732.322 (103) &   365.989 (00.014) &   49 (7) \\
$f_{C14}$ & 2735.515 (051) &   365.562 (00.007) &   99 (7) \\
$f_{C15}$ & 2771.443 (056) &   360.823 (00.007) &   88 (7) \\
$f_{C16}$ & 2805.022 (118) &   356.503 (00.015) &   42 (7) \\
$f_{C17}$ & 2865.190 (098) &   349.017 (00.012) &   50 (7) \\
$f_{C18}$ & 2916.403 (176) &   342.888 (00.021) &   28 (7) \\
$f_{D1}$ & 3529.254 (114) &   283.346 (00.009) &   45 (7) \\
$f_{D2}$ & 3531.300 (153) &   283.182 (00.012) &   34 (7) \\
$f_{D3}$ & 3557.908 (130) &   281.064 (00.010) &   38 (7) \\
$f_{D4}$ & 3578.434 (117) &   279.452 (00.009) &   43 (7) \\
$f_{D5}$ & 3639.159 (167) &   274.789 (00.013) &   30 (7) \\
$f_{D6}$ & 3674.105 (166) &   272.175 (00.012) &   30 (7) \\
$f_{D7}$$^\dag$ & 3712.247 (125) &   269.379 (00.008) &  131 (7) \\
$f_{D8}$$^\dag$ & 3712.369 (125) &   269.370 (00.008) &  -- (--) \\
$f_{D9}$ & 3778.092 (119) &   264.684 (00.008) &   42 (7) \\
$f_{E1}$ & 4400.021 (126) &   227.272 (00.007) &   39 (7) \\
$f_{E2}$ & 4531.554 (172) &   220.675 (00.008) &   29 (7) \\
\hline
\end{tabular}
\end{table}

The two pairs of frequencies marked with $^\dag$ are unresolved.
Prewhitening one frequency with
the measured amplitude leaves a significant residual peak, but fitting
two frequencies simultaneously produces possibly exaggerated amplitudes
and high errors on the frequencies. The amplitude given is therefore
only stated as the observed amplitude of the combined peak.

\end{document}